\documentclass[10pt,conference]{IEEEtran}
\IEEEoverridecommandlockouts
\usepackage[table]{xcolor}
\definecolor{oursblue}{RGB}{225,236,250}
\usepackage{booktabs} 
\usepackage{cite}
\usepackage{amsmath,amssymb,amsfonts}
\usepackage{algorithmic}
\usepackage{graphicx}
\usepackage{textcomp}
\usepackage{xcolor}
\usepackage{url}
\usepackage[hidelinks]{hyperref}

\usepackage{fontawesome5}
\usepackage[most]{tcolorbox}

\newcommand{\finding}[2]{\par\smallskip\noindent{\small #1}~\textbf{#2.}~}
\DeclareRobustCommand{\oursbox}[1]{{\setlength{\fboxsep}{1pt}\colorbox{oursblue}{\strut #1}}}

\newtcolorbox{answerbox}[1]{%
  enhanced, breakable=false,
  colback=gray!7, colframe=gray!55, boxrule=0.4pt, arc=2pt,
  left=4pt, right=4pt, top=3pt, bottom=3pt,
  before upper={{\faLightbulb[regular]~\textbf{Answer to RQ#1:}}~}}

 \newtcolorbox{challengebox}{enhanced, colback=gray!6, colframe=gray!55,
boxrule=0.4pt, arc=2pt, left=4pt, right=4pt, top=3pt, bottom=3pt}

\graphicspath{{figures/}}

\def\BibTeX{{\rm B\kern-.05em{\sc i\kern-.025em b}\kern-.08em
    T\kern-.1667em\lower.7ex\hbox{E}\kern-.125emX}}

\newcommand{\todo}[1]{}

\begin{document}

\title{CLEW: An Agentic LLM Approach for Log-Guided Root-Cause Code Localization}

\author{\IEEEauthorblockN{Wenji Bai, Muhammad Waseem, Jaakko Peltonen, Pekka Abrahamsson}
\IEEEauthorblockA{\textit{Tampere University}\\
Tampere, Finland\\
\{wenji.bai, muhammad.waseem, jaakko.peltonen, pekka.abrahamsson\}@tuni.fi}
}

\maketitle

\begin{abstract}
Failures in industrial software systems may involve multiple components and be difficult to reproduce outside the production environment. Root-cause code localization therefore often relies on issue descriptions and runtime logs. However, log evidence may be noisy or incomplete, and the code reporting a symptom may differ from the code containing the underlying defect. These conditions make it challenging to identify relevant log evidence and determine which functions could have caused the reported failure. To address this gap, we introduce CLEW, an agentic approach that combines rule-based log-to-code anchoring with LLM-guided backward code tracing. CLEW organizes selected observations into diagnostic states, guides source exploration from anchored symptom functions, and returns ranked candidate functions, an affected module, and an explanation comprising a proposed mechanism path and associated log events. For evaluation, we construct LogLoc, a benchmark comprising 192 real-world issue and fault-injection cases. Against three baselines, CLEW records the highest observed Acc@1 (77.3\%) and MRR (81.3\%) on LogLoc-Issues and matches the leading baselines at 97.6\% Acc@1 on LogLoc-Airflow. Averaged over all 192 cases, CLEW's wall-clock time is approximately 73\% lower than LocAgent's at a similar average inference cost. No aggregate degradation is observed under the tested perturbations, with modest changes in Acc@1 and MRR across the evaluated seeds and base models. These results suggest that CLEW can combine function-level localization with evidence links and runtime efficiency under the evaluated conditions.
\end{abstract}

\section{Introduction}\label{sec:intro}

In large industrial software systems, developers often investigate production failures using three inputs: an issue description, the production logs, and a snapshot of the deployed source code~\cite{yuan2010sherlog,wong2016survey}. A reproducible failure, or execution-derived information such as failing tests and coverage data, may not be available. Runtime logs record the system's observed behavior, while the source code provides context for interpreting it~\cite{yuan2010sherlog,zhang2017pensieve}. This paper addresses root-cause code localization: locating the defective function from these three inputs.

\begin{figure*}[t]
  \centering
  \includegraphics[width=\textwidth]{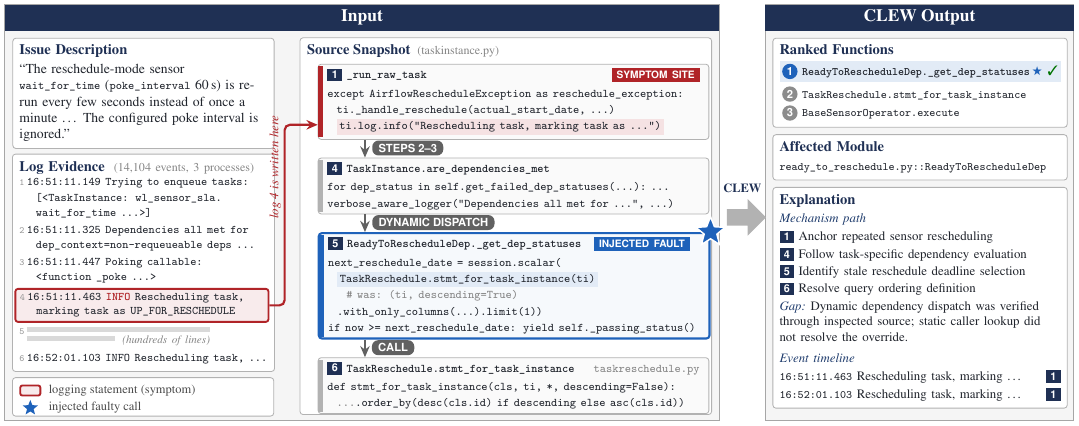}
  \caption{An illustrative fault-injection case from LogLoc-Airflow.}
  \label{fig:example}
\end{figure*}
However, using this evidence to locate the code responsible for a failure remains challenging. First, different logging statements may share the same message text~\cite{li2022duplicate}, so a log message matched on its text alone may correspond to several code locations, each of which becomes a candidate. Second, log evidence may be noisy and incomplete. Messages from concurrent requests and threads are interleaved~\cite{zhao2014lprof}, a single fault may trigger retries and downstream errors, and events that would explain the failure may never have been logged~\cite{yuan2012errlog}. As a result, the evidence needed to distinguish between candidates may be obscured by unrelated activity or missing altogether. Third, the code that reports a symptom is often not the code that causes it~\cite{avizienis2004taxonomy}. Downstream errors and stack frames direct attention to the code that detects or reports a symptom, whereas the defect may lie in code that executed earlier and whose effects propagated to the reporting location~\cite{wu2014crashlocator}. For example, in Fig.~\ref{fig:example}, an Apache Airflow task logs that a sensor is being rescheduled, while the injected fault lies in the dependency check that decides when the sensor may run again. Localization therefore often needs to trace backward from the code that reports the symptom to the code that caused it.

Prior work has explored several approaches to this challenge. Early fault localization techniques rank suspicious code using test outcomes and coverage~\cite{wong2016survey}, which are often unavailable because production failures cannot always be reproduced~\cite{yuan2010sherlog}. Log-based diagnosis avoids this dependence by linking logs to source code. SherLog and Pensieve infer execution paths from logs and code~\cite{yuan2010sherlog,zhang2017pensieve}, and COCA traces log messages back to their logging statements and uses an LLM to localize the root cause at the component level~\cite{coca}. LocAgent supports graph-guided code localization from issue reports~\cite{locagent}, while Agentless uses hierarchical code selection~\cite{agentless}. CLEW combines explicit log-to-code anchors with agent-driven backward source exploration and function-level candidate ranking. We examine whether this structured guidance improves function-level localization beyond an agent's own source exploration.



To address this challenge, we propose \textbf{CLEW}, an agentic approach for root-cause code localization through log-anchored backward tracing. First, CLEW anchors the log evidence to the code locations that may have produced it and selects the events relevant to the reported failure. By linking log observations to candidate source functions, these anchors provide starting points for investigation. Then, starting from the anchored functions, a code-tracing agent inspects the source code and traces upstream to find candidate root causes. Since an anchored function may only report a downstream symptom, tracing its code relationships helps identify other functions that could contain the defect. Finally, CLEW ranks the candidates, identifies the affected module, and returns an explanation consisting of the agent's proposed mechanism path linked to log observations. In Fig.~\ref{fig:example}, CLEW follows the code relationships from the function that logs the rescheduling to the reschedule dependency check, ranks the faulty function first, and links the proposed mechanism to the associated log events.



To evaluate CLEW, we construct \textbf{LogLoc}, a benchmark of 192 cases drawn from two complementary sources (Section~\ref{sec:benchmark}): real-world issue reports, which provide realistic diagnostic inputs, and fault-injection cases, which provide known fault locations and separate log bundles. LogLoc-Issues contains 150 real-world reports with embedded log evidence; its reference fix locations are non-test library functions modified by the reference fix and present in the source snapshot. These locations serve as localization proxies rather than independently confirmed production root causes. LogLoc-Airflow contains 42 controlled fault-injection cases with multi-process log bundles and uses the function containing each injected fault as its ground truth.

We compare CLEW with three baselines using the same base model. On LogLoc-Issues, CLEW achieves the highest observed Acc@1 of 77.3\%, 5.3 percentage points above LocAgent (72.0\%), although this difference is not statistically significant, and its Recall@5 is lower (73.2\% versus 75.6\%). On LogLoc-Airflow, CLEW matches LocAgent and Agentless at 97.6\% Acc@1. Averaged over all 192 cases, CLEW takes 60~s per case compared with 219~s for LocAgent, an approximately 73\% reduction, at a similar average inference cost (\$0.157 versus \$0.154 per case); per dataset, CLEW remains faster but costs more on LogLoc-Issues and less on LogLoc-Airflow. Under the two tested perturbation conditions on the applicable LogLoc-Issues subsets (inserting background log lines and removing intermediate traceback frames), aggregate Acc@1 does not decrease.

The contributions of this paper are as follows:
\begin{itemize}
  \item We introduce \textbf{CLEW}, whose central idea is to structure an LLM agent's source investigation using explicit log-to-code anchors. It uses anchored observations as diagnostic context and candidate starting points for backward exploration, and retains log-to-code links when presenting a proposed mechanism path.


   \item We construct \textbf{LogLoc}, a 192-case benchmark combining two complementary sources: 150 real-world reports with embedded log evidence, whose reference fix locations serve as localization proxies, and 42 controlled fault-injection cases with separate multi-process log bundles and known injected fault locations.

\item We evaluate CLEW against three baselines, obtaining the highest observed Acc@1 on LogLoc-Issues with lower Recall@5 than LocAgent, and matching the leading Acc@1 on LogLoc-Airflow. Averaged over the combined benchmark of 192 cases, CLEW has lower wall-clock time than LocAgent at similar inference cost.

\end{itemize}
\section{Approach}
\label{sec:approach}

\begin{figure*}[t]
  \centering
  \includegraphics[width=\textwidth]{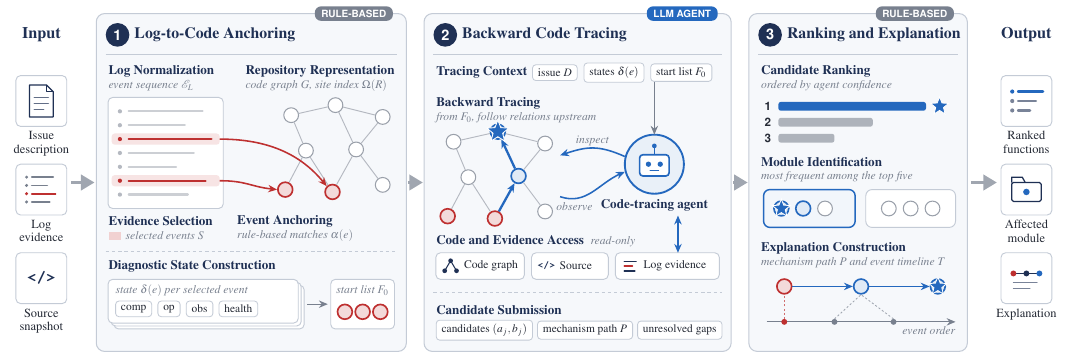}
  \caption{Overview of CLEW.}
  \label{fig:overview}
\end{figure*}
\subsection{Overview}
\label{sec:overview}
As illustrated in Fig.~\ref{fig:overview}, CLEW maps log events to source functions and organizes selected evidence into diagnostic states (Phase~1), then guides an agent's source investigation from anchored symptom functions toward potential upstream causes (Phase~2). It ranks the candidate functions, identifies the affected module, and links a proposed mechanism path to log observations (Phase~3). Implementation details are provided in the supplementary material (S1--S7).

\subsection{Task Formulation}
\label{sec:problem}
Following ICSE 2027 Industry Challenge~\#7~\cite{challenge7}, our objective is to localize root-cause code in software systems from issue descriptions and log evidence. The input consists of a natural-language issue description $D$, log evidence $L$ recording the observed behavior, provided as a separate log bundle or embedded in the description, and a snapshot $R$ of the source code as deployed when the failure occurred.

Our goal is to identify and rank the functions that may contain the root cause. We represent the result as a ranked list $C = [(c_1, s_1), \dots, (c_k, s_k)]$ ordered by decreasing score, where each candidate function $c_i$ is identified by its file path and qualified name and $s_i$ is its localization score. In addition, the task requires the affected module $m$ and an explanation $X$ describing proposed connections between the candidate code and the reported symptom, together with associated log observations.

\subsection{Phase 1: Log-to-Code Anchoring}
\label{sec:phase1}

\textbf{Repository representation.}
From the source snapshot $R$, CLEW constructs an observation-site index $\Omega(R)$ and a code graph $G$. The index records logging statements and their message templates, exception-raising statements and their exception types, and function locations for resolving stack frames. The graph captures statically resolved calls, attribute-based references, and method overrides. Both are constructed once per source snapshot.

\textbf{Log Normalization.}
CLEW begins log processing by normalizing $L$ into an event sequence $\mathcal{E}_L = (e_1, \dots, e_n)$. Events record available information about timestamps, components, message templates, correlation identifiers, severity, and execution context. Stack traces are parsed into frame records containing file paths, line numbers, and function names, with each record associated with an event.

Events in LogLoc-Issues retain their order of appearance in the report; for LogLoc-Airflow, events from the log bundle are merged by timestamp and consecutive repetitions are consolidated (S1). This event order is a processing convention and may differ from the actual order of occurrence.

\textbf{Event Anchoring.}
For each event $e \in \mathcal{E}_L$, CLEW constructs an ordered list $\alpha(e)$ of candidate source sites with rule-based confidence scores. Table~\ref{tab:anchor-rules} specifies the matching rules and their scores, which express match specificity and are not calibrated probabilities. The matched sites are kept as alternatives in $\alpha(e)$, ordered by decreasing score, up to a fixed cap per event (S1).

\begin{table}[t]
  \centering
  \caption{Log-to-code anchoring rules}
  \label{tab:anchor-rules}
  \footnotesize
  \setlength{\tabcolsep}{3pt}
  \begin{tabular}{c >{\raggedright\arraybackslash}p{0.62\columnwidth} c}
    \toprule
    \textbf{\#} & \textbf{Rule (evidence $\rightarrow$ site)} & \textbf{Score} \\
    \midrule
    1  & Stack frame, file and line resolve $\rightarrow$ enclosing function & 1.0 \\
    2  & Stack frame, file resolves, line does not $\rightarrow$ same-named function in file & 0.7 \\
    3  & Stack frame, line outside any function $\rightarrow$ module-level site & 0.4 \\
    4  & Stack frame, file does not resolve $\rightarrow$ function by name & 0.5 / 0.35$^{a}$ \\
    \addlinespace[2pt]
    5  & Exception type $\rightarrow$ raise sites of that type & 0.3--1.0$^{b}$ \\
    6  & Identifier in exception message $\rightarrow$ function by name & 0.6 / 0.45 / 0.3$^{b}$ \\
    \addlinespace[2pt]
    7  & Log line with source location $\rightarrow$ enclosing function & 1.0 \\
    8  & Log template, same level $\rightarrow$ logging sites & $J$ \\
    \addlinespace[2pt]
    9  & Error message $\rightarrow$ logging or raise sites & $0.3 + 0.5J$ \\
    10 & Identifier in error message $\rightarrow$ function by name & 0.4 / 0.3$^{a}$ \\
    \addlinespace[2pt]
    11 & Identifier in the issue description $\rightarrow$ function by name, not attached to an event & 0.4 / 0.25$^{a}$ \\
    \bottomrule
  \end{tabular}
  \textit{Note.} $J$ is the Jaccard similarity of template token sets. $^{a}$One / two to five functions share the name. $^{b}$The score depends on how specific the match is; the conditions are listed in S1.
\end{table}

\textbf{Evidence Selection.}
To identify evidence relevant to the reported problem, CLEW extracts a symptom profile from $D$ and the error-level events using fixed rules. The profile includes component names, code identifiers, source-file mentions, title keywords, and exception types where available.

CLEW scores each event with rule-based terms covering severity and lexical relevance to the symptom profile, using one procedure for log excerpts embedded in a report (LogLoc-Issues) and one for multi-file log bundles (LogLoc-Airflow); the terms and weights are listed in S1.

CLEW then forms the selected event set $S$ using a score threshold for excerpts and score-based ranking with template caps for bundles, retaining all error-level events in both settings; these settings were fixed before the evaluation cases were run (S1). Every event in $S$ receives a diagnostic state, and at most 60 states are rendered in the agent's initial context; the others remain retrievable through evidence lookup in Phase~2.

\textbf{Diagnostic State Construction.}
For each selected event $e \in S$, CLEW organizes the available information into a diagnostic state $\delta(e) = \langle \mathit{comp}, \mathit{op}, \mathit{obs}, \mathit{health} \rangle$, representing the component, operation, observation, and an observation-level health indication, respectively. All four fields are filled by fixed rules, not by the model (S1). The component and observation come from the normalized event. The health field marks a single observation as normal or abnormal and does not indicate the cause of the failure. The operation is the function of the event's highest-confidence anchor that resolves to a function; alternative anchors remain accessible through log-to-code lookup. Fields that cannot be grounded are marked \texttt{unknown}.

Finally, CLEW scans the diagnostic states in reverse event order and forms $F_0$ from the first three known operations  (S1).

\subsection{Phase 2: Backward Code Tracing}
\label{sec:phase2}

CLEW uses a code-tracing agent to inspect source functions and explore upstream structural relationships from $F_0$, with the issue description and diagnostic states providing context for the investigation.

\textbf{Tracing Context.}
The agent receives a fixed system prompt and a case-specific message containing the issue description $D$, the symptom profile, diagnostic states derived from the selected events $S$, and the start functions $F_0$. If no diagnostic state has a known operation, $F_0$ is empty and the agent is instructed to locate starting functions by searching the repository for identifiers and message fragments from the error-level events and the issue description (S2).

\textbf{Code and Evidence Access.}
To support the investigation, CLEW provides read-only access to the source snapshot $R$, code graph $G$, and observation-site index $\Omega(R)$. Graph navigation retrieves callers and callees. Source inspection retrieves function bodies, supports text searches across the repository, and locates possible assignments and reads by name-based text matching (S2), which is not a complete data-flow analysis. Evidence lookup retrieves an event's anchor alternatives $\alpha(e)$, lists logging and raise sites within a function, and finds sites whose templates match a given message.

\textbf{Backward Tracing and Candidate Submission.}
When $F_0$ is nonempty, the tracing instructions direct the agent to start from these functions and explore call, attribute-reference, and override relationships when investigating potential upstream causes. Here, ``backward'' denotes the prompted direction of investigation, from the anchored symptom functions toward their upstream dependencies; the tools themselves expose all three relationship types in either direction. These static connections do not by themselves establish causality. Tool-call and turn limits (Section~\ref{sec:setting}) bound the exploration, and CLEW requests a final submission as either limit is approached. Before proposing a candidate, the agent is instructed to inspect its source and describe its relationship to the symptom functions.

The investigation concludes through a submission tool. The submission contains a sequence of candidate--confidence pairs $A = [(a_1, b_1), \dots, (a_h, b_h)]$, where $a_j$ is a candidate function in $R$ and $b_j \in [0,1]$ is its agent-assigned confidence, the candidates' stated relationships to the symptom, and a single mechanism path $P = (\pi_1, \dots, \pi_q)$. The path presents the agent's account of how the reported symptom may arise, as one hypothesis for the whole submission; it need not include the first-ranked candidate, and candidates absent from it remain alternatives. Following the tracing instructions, its steps run from the symptom function toward a proposed cause. Each step gives a short description with the functions and log-event identifiers cited as evidence, and connections that lack sufficient code or log support are to be recorded as unresolved gaps (Fig.~\ref{fig:example} shows an example).

The implementation checks the submission (S2). A candidate identifier must resolve to a function in the snapshot index; unresolved and repeated candidates are discarded, and recovery and resubmission behavior is described in S2. Function and event references in the path are retained as submitted, without rejection or correction, and are examined by the grounding checks (Section~\ref{sec:analysis}); an omitted or empty path is accepted. Candidate checks and the tool-call and turn limits are enforced; source inspection, backward step ordering, and gap reporting are instructions to the agent.

\subsection{Phase 3: Ranking and Explanation}
\label{sec:phase3}

\textbf{Candidate Ranking.}
CLEW sorts the pairs in $A$ by decreasing agent-assigned confidence, without recomputing the scores, and retains at most $K=10$ candidates; because only the first eight submitted candidates are processed (S2), the effective limit is eight and $K$ is a nonbinding output cap. The resulting ranked list is $C = [(c_1, s_1), \dots, (c_k, s_k)]$, where $k \leq K$. If no candidate is obtained, the case receives an empty ranking $C = []$ and counts as a miss.

\textbf{Module Identification.}
The affected module is provided as an implemented output and is not evaluated separately; its unit is the class enclosing a function, or its file for module-level functions, following the module-level granularity of LocAgent~\cite{locagent}. CLEW selects as $m$ the most frequent module among the top five candidates (tie handling in S2).

\textbf{Explanation Construction.}
The explanation $X = \langle P, T \rangle$ combines the mechanism path $P$ submitted in Phase~2 with an event timeline $T$ containing log observations linked to the path or leading candidates.

CLEW constructs $T$ from the complete normalized event sequence $\mathcal{E}_L$ produced in Phase~1. An event is eligible when at least one of its anchors has confidence at or above a threshold $\theta$ and corresponds to a function cited in $P$ or among the top three candidates. Eligible events are kept in the Phase~1 event order, subject to the limits in Section~\ref{sec:setting}.

Each timeline entry links a log event to its associated function and the corresponding path step or candidate rank (fields in S2). An empty timeline is allowed and changes neither the ranking nor the submitted path.
\section{Experimental Setup}\label{sec:setup}

\subsection{Research Questions}
\label{sec:rqs}
We study three research questions (RQs).
\begin{itemize}
  \item \textbf{RQ1:} How effectively does CLEW  localize root-cause code from issue descriptions and log evidence?
  \item \textbf{RQ2:} How do the components of CLEW  contribute to its localization effectiveness?
  \item \textbf{RQ3:} How robust is CLEW  to noisy and incomplete log evidence?
\end{itemize}
RQ1 examines localization effectiveness and efficiency against three baselines across both sub-datasets and different distances between the anchored functions and the reference fix locations. RQ2 examines component contributions through ablation studies and checks the structural grounding of the explanations. RQ3 examines robustness to added log noise and missing stack frames. Agreement across seeds and sensitivity to the base model are reported as supplementary stability analyses.

\subsection{Benchmark}
\label{sec:benchmark}

In this paper, we build our evaluation dataset, LogLoc, from two sources: real-world issue reports and fault-injection cases.

\textbf{LogLoc-Issues.}
This sub-dataset contains 150 Python issue cases from 25 repositories, comprising 47 cases from SWE-bench Verified~\cite{jimenez2024swebench} and 103 from the Python subset of MULocBench~\cite{zhang2025benchmark}. Selection was rule-based and exhaustive. From the 500 SWE-bench Verified instances and the 842 Python MULocBench issues, we kept the 274 cases with log or error evidence and valid reference fix locations, after removing duplicate fixing pull requests, and partitioned them by repository to obtain the 150 cases used here. The rules and counts of each step are given in S3. Each case includes an issue description, accompanying log or error evidence, and the repository snapshot at the base commit. For localization evaluation, reference fix locations are non-test library functions modified by the fix and present at the base commit, identified through patch-to-function mapping or MULocBench's function-level annotations; functions added by the fix are therefore excluded, and deleted or renamed functions are identified by their name at the base commit (S3).

\textbf{LogLoc-Airflow.}
This sub-dataset contains 42 cases from 21 single-function injection points in Apache Airflow~2.10.5. Each point runs under two workload configurations, yielding two related cases treated as one analysis cluster. The points were selected from a catalog of 30 scheduling and execution mutations (e.g., removed state transitions, inverted conditions, and omitted calls), after excluding one unobservable mutation and partitioning the rest by injection point. The configurations run each point's designated workloads or all five workloads (S4). All 42 cases satisfied their symptom rules; none of the rules held on the fault-free baseline, and no additional runs were excluded. One rule was narrowed to the affected task after the second configuration, without changing first-configuration outcomes. Each case provides a symptom description, logs, and the deployed source snapshot; the mutated function is the localization reference. Descriptions were prepared from the catalog before inspecting runs and omit the faulty function, its file, and mutation details, while permitting error-message quotations. Injection-revealing harness identifiers are masked in all method inputs.

\subsection{Metrics}
\label{sec:metrics}
For each case $q$ in the evaluated set $Q$, $Y_q$ is the set of reference functions, $C_q$ is the ranked list returned by a method, and $C_q^{(K)}$ contains its first $K$ entries. A missing or invalid output is treated as $C_q = []$. Functions are matched by file path and qualified name, a repeated function counts once, and rankings shorter than $K$ are evaluated as returned; $Y_q$ is never empty.

Acc@$K$ is the proportion of cases with at least one reference function among the top $K$ predictions, 
$\frac{1}{|Q|}\sum_{q \in Q} \mathbf{1} [Y_q \cap C_q^{(K)} \neq \emptyset]$. 
We use Acc@1 as a primary measure and additionally report Acc@3 and Acc@5. Recall@$K$ is the fraction of a case's reference functions among the top $K$ predictions, averaged across cases, $\frac{1}{|Q|}\sum_{q \in Q} |Y_q \cap C_q^{(K)}| / |Y_q|$. We report Recall@5; on LogLoc-Issues, a case can have several reference fix locations. MRR is the mean reciprocal rank of the first reference function in $C_q$, $\frac{1}{|Q|}\sum_{q \in Q} 1/\mathrm{rank}_q$, with a reciprocal of zero if none is retrieved.

\subsection{Baselines}
\label{sec:baselines}

We compare CLEW with three baselines representing graph-guided exploration, hierarchical localization, and retrieval-based localization.

\begin{itemize}
 \item Base: This baseline ranks all non-test functions with BM25~\cite{robertson2009bm25} over their source code, using the input text as the query, and passes the top 30 functions to the LLM in a single prompt. The LLM then returns up to ten ranked candidate locations, without any further interaction with the repository.

  \item LocAgent~\cite{locagent}: This approach represents a repository as a heterogeneous code graph containing code entities and their structural relationships. An LLM-based agent navigates this graph through tools to identify code relevant to the reported issue. We use the authors' implementation (commit \texttt{4935b55}).

  \item Agentless~\cite{agentless}: This approach uses a hierarchical localization pipeline that progressively narrows the search from repository files to relevant classes and functions, followed by fine-grained code locations. We use the localization stage of the authors' implementation (commit \texttt{5ce5888}).
\end{itemize}

All methods access the same source snapshot. The three baselines receive identical text per case: the issue title and description, capped at 12,000 characters, which on LogLoc-Issues already embed the log evidence. On LogLoc-Airflow, the text also includes a raw log excerpt of up to 48,000 characters, a fixed setting applied to all cases and baselines. The excerpt concatenates the scheduler, DAG-processor, per-file processor, and task-attempt logs in path order, giving each file an equal share and keeping the beginning and end of over-limit text. By contrast, CLEW processes the full report in Phase~1, while its agent and its agent-based variant each receive at most 3,500 characters of report text. Truncation limits, prompts, and adaptations are given in S5.

\subsection{Ablation Variants}
\label{sec:variants}

For RQ2, we evaluate two variants on the same cases and with CLEW's settings: (1) CLEW w/o structured anchoring runs the same agent without diagnostic states, the tracing start list, or log-to-code lookup; and (2) Anchor floor ranks functions anchored to selected events in reverse event order and by decreasing anchor confidence within each event, without an LLM. On LogLoc-Issues, the first variant reads the same report text as CLEW. On LogLoc-Airflow, it reads the baselines' raw log excerpt, whereas CLEW selects and anchors observations from the full bundle; this comparison therefore assesses evidence selection and structured anchoring together, without isolating anchoring. A description-only Airflow configuration is reported as a reference. The anchor-floor comparison removes the agent as a whole and does not isolate backward tracing.

\subsection{Experimental Settings}
\label{sec:setting}

All methods use the fixed model version \texttt{openai/gpt-6.1-sol}, accessed through OpenRouter with medium reasoning effort, temperature~0, and a fixed seed (seed~0). These settings and the source snapshot are shared by all methods, while interaction limits and retry behavior follow each method's implementation. To examine run-to-run stability, we repeat the CLEW runs with a second seed, and to test model sensitivity, we evaluate CLEW with \texttt{google/gemini-4-argon}; both evaluations cover the full LogLoc dataset with all other settings unchanged.

CLEW and its agent-based variant use at most 40 tool calls and 30 model turns, plus up to three forced-submission turns, with at most 2,048 output tokens per turn and 4,096 for forced submissions; the anchor floor makes no model calls. LocAgent uses a limit of 25 model calls and 40 tool calls, while Agentless and Base follow fixed pipelines. Requests failing with rate-limit, server, or transport errors are attempted up to three times for CLEW, its variant, and Base, while LocAgent and Agentless keep their own retry logic. Beyond CLEW's submission recovery (Section~\ref{sec:phase2}), Base falls back to its retrieval order when its output cannot be parsed, and a case without a valid ranking counts as a miss for every method. For explanation construction, CLEW uses $\theta = 0.3$ and retains at most 40 timeline events and three per function--template pair.

Inference cost is computed from the input and output tokens of all completed model calls of a case at the provider's token prices. Wall-clock time runs from the start of a case to its final ranking and includes log anchoring and the construction of each method's code index or graph. Token prices, timeouts, the runtime environment, concurrency, caching, and execution dates are documented in S5.

\subsection{Data Analysis}
\label{sec:analysis}

We compare methods using paired case-level differences, with cases weighted equally. For Acc@1 and MRR in RQ1 and RQ2, we perform two-sided sign-flip permutation tests at the cluster level: per-case differences are summed within each cluster, and the signs of the cluster sums are permuted (exact enumeration up to 20 non-zero clusters, otherwise 200,000 random permutations). We apply Holm correction separately to CLEW's comparisons with the baselines (RQ1) and with the ablation variants (RQ2) within each sub-dataset and metric. We compute 95\% percentile confidence intervals using 10,000 cluster-bootstrap resamples, with repositories as clusters on LogLoc-Issues and injection points as clusters on LogLoc-Airflow. All differences, 95\% confidence intervals, and Holm-adjusted $p$-values for RQ1 and RQ2 are listed in S7.

Stratified localization results, cost and time per case, and automatic explanation-grounding checks are reported descriptively. Distance is the shortest path from any function anchored by a log event with confidence at least 0.3 (rules 1--10 in Table~\ref{tab:anchor-rules}) to any reference fix location over call, attribute-reference, and override edges in either direction. The search is bounded at eight hops, and cases are grouped into distance~0 (a reference fix location is itself anchored by a log event), distance~$\geq 1$ (a path is found within the bound), and no path found within eight hops, which includes cases without log-derived anchors and does not imply that the functions are unreachable. The grounding checks are computed by the evaluation scripts after generation and assess structural links, not causal correctness. Over all cases, they report whether the path starts at an anchored function and cites at least one existing log event, and whether the timeline is non-empty and adds events not cited by the path; an empty path fails both path checks but not the timeline checks. For the link check, steps without a resolvable function are removed, and consecutive remaining functions count as linked when they are identical or at most two hops apart over the same edges, with the linked share pooled per sub-dataset (S6).

For RQ3, we compare perturbed and original results on the same applicable cases. Perturbations are generated once with a fixed seed. \emph{Noise} randomly inserts background log lines from other issues in the same repository, excluding lines that mention the case's reference locations, matching the number of non-empty lines in the original evidence block. \emph{Incomplete} removes half of the intermediate traceback frames (rounded up) and their source lines, preserving the first and last frames and the exception line. All other evidence is preserved. Noise applies to cases with an evidence block and enough background lines, and Incomplete to cases with a removable frame (S6). The absolute difference between two unperturbed seed runs serves as a descriptive reference; we report bootstrap confidence intervals for changes in Acc@1.

Stability is analyzed through metric differences and Top-1 correctness agreement across seeds and base models: the proportion of cases on which both runs rank a reference function first or both fail to do so, with empty or invalid predictions counted as misses. This measures agreement in correctness.
\section{Results}\label{sec:results}

\subsection{Localization Effectiveness (RQ1)}
\label{sec:rq1}

Table~\ref{tab:rq1} reports the function-level results and the average cost and time per case of each method, and Table~\ref{tab:strata} reports Acc@1 on LogLoc-Issues by the minimum code-graph distance from log-anchored functions to reference fix locations.

\begin{table}[t]
  \centering
  \caption{Localization results (\%), cost (\$), and time (s) on LogLoc.}
  \label{tab:rq1}
  \footnotesize
  \setlength{\tabcolsep}{2.6pt}
  \begin{tabular}{lccccc|cc}
    \toprule
    \textbf{Method} & \textbf{Acc@1} & \textbf{Acc@3} & \textbf{Acc@5} & \textbf{Rec@5} & \textbf{MRR} & \textbf{Cost} & \textbf{Time} \\
    \midrule
    \rowcolor{gray!12}
    \multicolumn{8}{c}{\textbf{LogLoc-Issues} ($n = 150$)} \\
    Base & 60.7$^{\dagger}$ & 64.7 & 65.3 & 52.6 & 62.8$^{\dagger}$ & 0.027 & 10 \\
    Agentless   & 70.0 & 79.3 & 79.3 & 64.7 & 74.3$^{\dagger}$ & 0.031 & 25 \\
    LocAgent    & 72.0 & 83.3 & 86.0 & \textbf{75.6} & 78.6 & 0.103 & 256 \\
    \rowcolor{oursblue}
    CLEW        & \textbf{77.3} & \textbf{85.3} & \textbf{86.7} & 73.2 & \textbf{81.3} & 0.125 & 57 \\
    \midrule
    \rowcolor{gray!12}
    \multicolumn{8}{c}{\textbf{LogLoc-Airflow} ($n = 42$)} \\
    Base & 31.0$^{\dagger}$ & 31.0 & 31.0 & 31.0 & 31.0$^{\dagger}$ & 0.060 & 22 \\
    Agentless   & \textbf{97.6} & 97.6 & 97.6 & 97.6 & 97.6 & 0.171 & 31 \\
    LocAgent    & \textbf{97.6} & \textbf{100.0} & \textbf{100.0} & \textbf{100.0} & \textbf{98.8} & 0.337 & 87 \\
    \rowcolor{oursblue}
    CLEW        & \textbf{97.6} & \textbf{100.0} & \textbf{100.0} & \textbf{100.0} & \textbf{98.8} & 0.269 & 72 \\
    \bottomrule
  \end{tabular}
  \par\smallskip
  \begin{minipage}{\columnwidth}\scriptsize
  \textit{Note.} Seed~0. $^{\dagger}$ marks baseline values significantly lower than CLEW (cluster-level sign-flip permutation test, Holm-adjusted, $p < .05$; tested for Acc@1 and MRR). \textbf{Bold} marks the best localization value per column. \oursbox{Blue background} indicates the results of CLEW. Time covers each case through its final ranking, including index or graph construction. Cost and time are case-level means; combined means weight all 192 cases equally.
  \end{minipage}
\end{table}

\begin{table}[t]
  \centering
  \caption{Acc@1 (\%) on LogLoc-Issues by distance from log-anchored functions to reference fix locations.}
  \label{tab:strata}
 \footnotesize
  \setlength{\tabcolsep}{4pt}
  \begin{tabular}{l r cccc}
    \toprule
    \textbf{Distance} & $\boldsymbol{n}$ & \textbf{Base} & \textbf{Agentless} & \textbf{LocAgent} & \textbf{CLEW} \\
    \midrule
    0 (anchored)            & 74 & 78.4 & 81.1 & 82.4 & \textbf{91.9} \\
    $\geq 1$ (path found)   & 45 & 35.6 & 60.0 & 55.6 & \textbf{62.2} \\
    no path found (8 hops)  & 31 & 54.8 & 58.1 & \textbf{71.0} & 64.5 \\
    \bottomrule
  \end{tabular}
\end{table}

\finding{\faChartBar}{Effectiveness on LogLoc-Issues}
On LogLoc-Issues, CLEW achieves the highest observed Acc@1 of 77.3\%, which is 5.3, 7.3, and 16.7 percentage points above LocAgent, Agentless, and Base, respectively. After Holm adjustment, the difference is significant against Base but not against LocAgent or Agentless (S7). CLEW also has the highest observed Acc@3, Acc@5, and MRR, while its Recall@5 is 2.4 percentage points lower than LocAgent's. Among the 128 cases in which both methods include a reference fix location in their returned rankings, CLEW places the first such function at an average rank of 1.16, compared with 1.39 for LocAgent (1.10 versus 1.17 in the 124 cases where both do so within their top five predictions). Thus, CLEW provides an earlier entry point on average in these shared successful cases, while LocAgent covers a larger proportion of reference fix locations within its top five predictions.

\finding{\faServer}{Effectiveness on LogLoc-Airflow}
On LogLoc-Airflow, CLEW, LocAgent, and Agentless each place the mutated function first in 41 of the 42 cases (97.6\% Acc@1), while Base reaches 31.0\%. CLEW and LocAgent assign the same rank to the mutated function in all 42 cases, and CLEW and Agentless each rank it higher than the other in one case. On this sub-dataset, CLEW therefore performs comparably to the leading evaluated baselines.

\finding{\faSitemap}{Effect of the distance from log anchors}
As shown in Table~\ref{tab:strata}, when a reference fix location is directly anchored by a log event (74 cases), CLEW achieves an Acc@1 of 91.9\%, 9.5 percentage points above LocAgent's 82.4\%. When it is connected to a log-anchored function through one or more code-graph hops (45 cases), CLEW achieves 62.2\%, the highest observed among the compared methods. In the remaining 31 cases, for which no path was found within eight hops, CLEW achieves 64.5\%, while LocAgent reaches 71.0\%.

\finding{\faCoins}{Efficiency in cost and time}
Averaged over all 192 cases, CLEW and LocAgent incur similar inference costs (\$0.157 and \$0.154 per case), while CLEW takes 60~s per case compared with 219~s for LocAgent, a reduction of approximately 73\%. Per sub-dataset (Table~\ref{tab:rq1}), CLEW costs more than LocAgent on LogLoc-Issues and less on LogLoc-Airflow, and takes less time on both. Agentless and Base, which do not explore the repository iteratively, have lower costs and shorter runtimes on both sub-datasets.

\begin{answerbox}{1}
CLEW achieves the highest observed Acc@1, Acc@3, Acc@5, and MRR among the evaluated methods on LogLoc-Issues, although its Acc@1 difference from LocAgent is not significant and its Recall@5 is lower than LocAgent's. On LogLoc-Airflow, CLEW matches LocAgent and Agentless at 97.6\% Acc@1. Averaged over all 192 cases, CLEW's inference cost is similar to LocAgent's and its wall-clock time per case is approximately 73\% lower; on each sub-dataset, CLEW remains faster, but its cost is higher on LogLoc-Issues and lower on LogLoc-Airflow.
\end{answerbox}

\subsection{Component Contribution (RQ2)}
\label{sec:rq2}

To answer RQ2, we compare CLEW with the variants defined in Section~\ref{sec:variants} on the same cases (Table~\ref{tab:rq2}). We additionally apply automatic grounding checks to CLEW's mechanism paths and event timelines (Table~\ref{tab:grounding}).

\begin{table}[t]
  \centering
  \caption{Ablation results (\%) on LogLoc.}
  \label{tab:rq2}
  \footnotesize
  \setlength{\tabcolsep}{4pt}
  \begin{tabular}{@{}lccccc@{}}
    \toprule
    \textbf{Configuration} & \textbf{Acc@1} & \textbf{Acc@3} & \textbf{Acc@5} & \textbf{Recall@5} & \textbf{MRR} \\
    \midrule
    \rowcolor{gray!12}
    \multicolumn{6}{c}{\textbf{LogLoc-Issues} ($n = 150$)} \\
    Anchor floor             & 20.0$^{\dagger}$ & 27.3 & 31.3 & 26.4 & 24.5$^{\dagger}$ \\ 
    w/o struct.\ anchoring   & 75.3 & \textbf{86.0} & \textbf{88.0} & 72.8 & 80.7 \\
    \rowcolor{oursblue}
    CLEW                     & \textbf{77.3} & 85.3 & 86.7 & \textbf{73.2} & \textbf{81.3} \\
    \midrule
    \rowcolor{gray!12}
    \multicolumn{6}{c}{\textbf{LogLoc-Airflow} ($n = 42$)} \\
    Anchor floor             & 4.8$^{\dagger}$ & 14.3 & 14.3 & 14.3 & 9.1$^{\dagger}$ \\
    w/o struct.\ anchoring   & 92.9 & 97.6 & 97.6 & 97.6 & 95.2 \\
    description only   & 92.9 & \textbf{100.0} & \textbf{100.0} & \textbf{100.0} & 96.4 \\
    \rowcolor{oursblue}
    CLEW                     & \textbf{97.6} & \textbf{100.0} & \textbf{100.0} & \textbf{100.0} & \textbf{98.8} \\
    \bottomrule
  \end{tabular}
  \par\smallskip
  \begin{minipage}{\columnwidth}\scriptsize
   \textit{Note.} $^{\dagger}$ marks variant values significantly lower than CLEW (same test as in Table~\ref{tab:rq1}). Cluster-bootstrap 95\% CIs of the Acc@1 difference (CLEW minus variant, in percentage points) are [47.6, 63.8] for the anchor floor and [$-$0.8, 4.8] for w/o struct.\ anchoring on LogLoc-Issues, and [85.7, 100.0] for the anchor floor and [0.0, 11.9] for both w/o struct. anchoring and description only on LogLoc-Airflow.
  \end{minipage}
\end{table}

\finding{\faCodeBranch}{Contribution of the code-tracing agent}
As shown in Table~\ref{tab:rq2}, CLEW's Acc@1 is significantly higher than that of the anchor floor on both sub-datasets (77.3\% versus 20.0\% on LogLoc-Issues and 97.6\% versus 4.8\% on LogLoc-Airflow). The code-tracing agent thus provides useful context beyond the functions ranked by the anchor floor. Because this comparison removes the agent as a whole, it does not isolate backward tracing from the agent's other capabilities, such as source inspection and repository search.

\finding{\faAnchor}{Contribution of structured anchoring}
Compared with the variant without structured anchoring, CLEW's Acc@1 is 2.0 percentage points higher on LogLoc-Issues (77.3\% versus 75.3\%) and 4.8 points higher on LogLoc-Airflow (97.6\% versus 92.9\%). Neither difference is statistically significant (Table~\ref{tab:rq2}). On LogLoc-Issues, the variant achieves slightly higher Acc@3 and Acc@5 than CLEW. On LogLoc-Airflow, the two configurations differ at only two of the 21 injection points, both favoring CLEW, and the variant has a higher inference cost (\$0.430 versus \$0.269 per case). The Airflow comparison evaluates evidence selection and structured anchoring together; its observed difference cannot be attributed to anchoring alone.

\begin{table}[t]
  \centering
  \caption{Automatic structural grounding checks of CLEW's explanations (mechanism path and event timeline) on LogLoc.}
  \label{tab:grounding}
  \footnotesize
  \setlength{\tabcolsep}{3pt}
  \begin{tabular*}{\columnwidth}{@{\extracolsep{\fill}}>{\raggedright\arraybackslash}p{0.48\columnwidth} r r@{}}
    \toprule
    \textbf{Check (\%)} & \textbf{LogLoc-Issues} & \textbf{LogLoc-Airflow} \\
                        & ($n = 150$) & ($n = 42$) \\
    \midrule
    \rowcolor{gray!12}
    \multicolumn{3}{@{}c@{}}{\textbf{Mechanism path}} \\
    Consecutive function pairs linked \mbox{($\leq 2$ hops)} & 71.8  & 92.4 \\
    Path starts at an anchored function       & 86.0  & 100.0 \\
    Path cites $\geq 1$ event                 & 97.3  & 69.0 \\
    \midrule
    \rowcolor{gray!12}
    \multicolumn{3}{@{}c@{}}{\textbf{Event timeline}} \\
    Timeline non-empty                        & 86.7  & 100.0 \\
    Timeline adds uncited events              & 16.7  & 100.0 \\
    \bottomrule
  \end{tabular*}
\end{table}

\finding{\faClipboardCheck}{Explanation Grounding}
Table~\ref{tab:grounding} summarizes the automatic checks defined in Section~\ref{sec:analysis}. On LogLoc-Issues and LogLoc-Airflow, 86.0\% and 100\% of mechanism paths start at an anchored function. For the link check, consecutive function pairs are formed after removing steps without resolvable functions (their counts are given in S6), rather than from all adjacent steps of the original paths. Of these pairs, 585 of 815 (71.8\%) and 257 of 278 (92.4\%) are linked within two hops in the code graph. The path cites at least one log event in 97.3\% and 69.0\% of cases. The timeline is non-empty in 86.7\% and 100\% of cases and adds events not cited by the path in 16.7\% and 100\%; in 13 LogLoc-Airflow cases, it provides log observations when the path itself cites none. These checks assess structural grounding, not the causal correctness of the proposed mechanisms.

\begin{answerbox}{2}
CLEW achieves significantly higher Acc@1 than the anchor floor on both sub-datasets. Compared with the variant without structured anchoring, metric changes are mixed and Acc@1 gains are not significant; on LogLoc-Airflow, this comparison evaluates evidence selection and anchoring jointly. Structural checks show that most proposed paths start at anchored functions, while the timeline adds log observations not cited by the path in all LogLoc-Airflow cases.
\end{answerbox}

\subsection{Robustness to Degraded Log Evidence (RQ3)}
\label{sec:rq3}

To answer RQ3, we evaluate CLEW on LogLoc-Issues under the two perturbation conditions defined in Section~\ref{sec:analysis}, which apply to 131 (Noise) and 84 (Incomplete) cases. Each perturbed result is compared with the original result on the same applicable cases, so the \emph{Original} rows of the two conditions in Table~\ref{tab:rq3} cover different case subsets. As a descriptive reference for run-to-run variability, we report the absolute difference between two unperturbed seed runs on the same cases as Seed variation~\cite{madaan2024variance,jaggedfrontier2026}.

\begin{table}[t]
  \centering
  \caption{CLEW under degraded log evidence (\%) on LogLoc-Issues.}
  \label{tab:rq3}
  \footnotesize
  \setlength{\tabcolsep}{4pt}
  \begin{tabular}{@{}l ccc@{}}
    \toprule
    \textbf{Run} & \textbf{Acc@1} & \textbf{Recall@5} & \textbf{MRR} \\
    \midrule
    \rowcolor{gray!12}
    \multicolumn{4}{@{}c@{}}{\textbf{Noise} ($n = 131$)} \\
    Original (seed 0)              & 78.6   & 72.9   & 82.1 \\
    Perturbed (seed 0)             & 80.2   & 74.0   & 83.4 \\
    $\Delta$ (perturbed $-$ original) & $+$1.5 & $+$1.1 & $+$1.3 \\
    Seed variation                 & 1.5    & 0.8    & 0.4 \\
    \midrule
    \rowcolor{gray!12}
    \multicolumn{4}{@{}c@{}}{\textbf{Incomplete} ($n = 84$)} \\
    Original (seed 0)              & 72.6   & 67.0   & 76.7 \\
    Perturbed (seed 0)             & 72.6   & 69.3   & 77.9 \\
    $\Delta$ (perturbed $-$ original) & 0.0 & $+$2.3 & $+$1.2 \\
    Seed variation                 & 2.4    & 0.9    & 0.9 \\
    \bottomrule
  \end{tabular}
  \par\smallskip
  \begin{minipage}{\columnwidth}\scriptsize
    \textit{Note.} Cluster-bootstrap 95\% CIs of $\Delta$Acc@1 are [$-$1.5, 4.6] under Noise and [$-$5.3, 6.2] under Incomplete.
  \end{minipage}
\end{table}

\finding{\faRandom}{Performance under noise and incomplete tracebacks}
Under Noise, CLEW's Acc@1 is 80.2\%, compared with 78.6\% on the original reports, a difference equal to the seed variation; Recall@5 and MRR also do not decrease. Under Incomplete, Acc@1 remains unchanged at 72.6\%, while Recall@5 and MRR increase by more than their seed variation. Thus, no aggregate degradation is observed under the tested conditions. However, the 95\% confidence intervals of the Acc@1 changes include negative values under both conditions ($[-1.5, 4.6]$ and $[-5.3, 6.2]$ percentage points for Noise and Incomplete, respectively).

\begin{answerbox}{3}
Under the tested perturbations, we observe no aggregate degradation in Acc@1, Recall@5, or MRR, although the 95\% confidence intervals of the Acc@1 changes include negative values. The Acc@1 changes are no larger than the corresponding differences between the two unperturbed seed runs, which provide only a limited descriptive reference.
\end{answerbox}

\subsection{Stability Across Seeds and Base Models}
\label{sec:stability}

\finding{\faSync}{Variation across seeds and models}
Table~\ref{tab:stability} compares CLEW across two seeds with GPT-6.1~Sol and across two base models with seed~0. Across the two seeds, Acc@1 differs by 1.3 percentage points on LogLoc-Issues and 2.4 on LogLoc-Airflow, and the Top-1 correctness agreement is 98.7\% and 92.9\%. With Gemini~4~Argon, CLEW obtains an Acc@1 of 76.0\% and 92.9\%, compared with 77.3\% and 97.6\% using GPT-6.1~Sol. Top-1 correctness agreement with GPT-6.1~Sol (seed~0) is 90.7\% on LogLoc-Issues and 95.2\% on LogLoc-Airflow. The largest observed model difference is in Recall@5 on LogLoc-Issues (78.3\% versus 73.2\%).

\begin{table}[t]
  \centering
  \caption{CLEW's localization results across seeds and base models (\%).}
  \label{tab:stability}
  \footnotesize
  \setlength{\tabcolsep}{4pt}
  \begin{tabular}{@{}l cccc@{}}
    \toprule
    \textbf{Model (seed)} & \textbf{Acc@1} & \textbf{Recall@5} & \textbf{MRR} & \textbf{\begin{tabular}[c]{@{}c@{}}Top-1 correctness\\agreement\end{tabular}} \\
    \midrule
    \rowcolor{gray!12}
    \multicolumn{5}{@{}c@{}}{\textbf{LogLoc-Issues} ($n = 150$)} \\
    GPT-6.1 Sol (0)    & 77.3 & 73.2 & 81.3 & ---  \\
    GPT-6.1 Sol (1)    & 76.0 & 73.9 & 80.9 & 98.7 \\
    Gemini 4 Argon (0) & 76.0 & 78.3 & 81.8 & 90.7 \\
    \midrule
    \rowcolor{gray!12}
    \multicolumn{5}{@{}c@{}}{\textbf{LogLoc-Airflow} ($n = 42$)} \\
    GPT-6.1 Sol (0)    & 97.6 & 100.0 & 98.8 & ---  \\
    GPT-6.1 Sol (1)    & 95.2 & 100.0 & 97.6 & 92.9 \\
    Gemini 4 Argon (0) & 92.9 & 97.6 & 95.2 & 95.2 \\
    \bottomrule
  \end{tabular}
    \par\smallskip
  \begin{minipage}{\columnwidth}\scriptsize
    \textit{Note.} Top-1 correctness agreement is measured against GPT-6.1~Sol (seed~0).
  \end{minipage}
\end{table}
\section{Discussion}\label{sec:discussion}

\textbf{Log-to-Code Anchoring Structures the Investigation Context.} Structured log-to-code anchoring provides the investigation with context and traceability. In RQ2, removing the code-tracing agent reduces performance substantially, whereas removing structured anchoring produces only small changes in the localization metrics, with no consistent direction. The current evidence therefore does not establish structured anchoring as a consistent accuracy improvement over the agent's own exploration. The agent may recover much of this context from the logs embedded in the LogLoc-Issues reports. On LogLoc-Airflow, the comparison evaluates evidence selection and structured anchoring together.

At the same time, anchoring adds a traceability feature that localization rank does not capture. CLEW retains the correspondence between log observations and code locations explicitly in its output: most mechanism paths originate at an anchored function, and the event timeline associates the relevant log events with the candidate functions. The evaluation checks these links only structurally: it does not show that they are causally correct or that they reduce engineers' investigation effort. They may still give engineers concrete leads for further inspection.

\textbf{The Relationship Between Log Anchors and Reference Fix Locations.} The RQ1 results suggest that log anchors may provide useful starting points for code exploration when they directly identify a reference fix location or are structurally connected to one. CLEW has the highest observed Acc@1 in these groups, while LocAgent has a higher Acc@1 where no path is found within eight hops. LocAgent derives search keywords from the issue text for graph-guided exploration~\cite{locagent}, whereas CLEW additionally starts from explicit log-to-code anchors. The pattern thus characterizes log-guided exploration as most competitive when the logs point at or near the reference fix locations, and suggests that broader issue-driven search may be useful when anchors offer limited guidance. This is a descriptive association: the groups contain different cases and may differ in difficulty. A missing path may reflect the search bound or limitations of the static graph, which may not fully represent dynamic dispatch, callbacks, or configuration-driven calls; it does not imply that the functions are unrelated at runtime.

\textbf{Implications for Research and Practice.} For researchers, the results suggest assessing both localization accuracy and the traceability of log-to-code links. Comparisons that vary log availability while keeping the description and source code fixed could help examine the contribution of log evidence. Overall accuracy and case-level consistency answer different evaluation questions: on LogLoc-Issues, the two base models achieve similar Acc@1 (77.3\% and 76.0\%) but 90.7\% Top-1 correctness agreement, so they succeed on partly different cases. Evaluations could therefore report both aggregate metrics and case-level correctness agreement across repeated runs~\cite{madaan2024variance,jaggedfrontier2026}. The observed robustness and stability should be interpreted within the tested inputs, runs, and models.

For practitioners, CLEW's ranked functions, mechanism path, and associated log events may support initial issue triage. Engineers can inspect candidate code alongside the linked log observations when checking a proposed mechanism. On LogLoc-Airflow, CLEW averages \$0.269 in inference cost per case versus \$0.430 for the raw-excerpt variant, without lower observed Acc@1. This configuration-level comparison does not isolate the contributions of evidence selection, context size, and anchoring to the cost difference, but suggests that preprocessing log bundles before agentic investigation may be worth considering where inference cost matters. Issue-driven repository search may provide a complementary starting point when the available log evidence offers limited guidance.
\section{Related Work}\label{sec:related}
 
\textbf{Fault and bug localization.} Spectrum-based techniques rank code using passing and failing test coverage~\cite{wong2016survey,abreu2007ochiai}, and BugLocator ranks source files from bug reports~\cite{zhou2012buglocator}. Pathidea reconstructs execution paths from report logs to improve file rankings~\cite{chen2022pathidea}, and CrashLocator ranks functions from crash stacks, including functions absent from the stack~\cite{wu2014crashlocator}. LLM-based techniques learn fault scores without coverage~\cite{yang2024llmao} or navigate repositories from a failing test~\cite{kang2024autofl}. CLEW focuses on function-level localization from issue descriptions, runtime logs, and deployed source snapshots. Unlike spectrum-based techniques and test-driven approaches such as AutoFL~\cite{abreu2007ochiai,kang2024autofl}, it does not require a failing test.
 
\textbf{Log analysis and root cause analysis.} Log analysis covers template extraction and anomaly detection~\cite{he2017drain,du2017deeplog}, and root cause analysis for cloud systems diagnoses faulty services from logs, metrics, and traces~\cite{soldani2022survey}, increasingly with LLMs~\cite{chen2024rcacopilot,roy2024exploring,jiang2025logpilot}. Closer to CLEW, SherLog and Pensieve infer failed executions or reproduction steps from logs and code~\cite{yuan2010sherlog,zhang2017pensieve}, and COCA combines logging-source retrieval and execution-path reconstruction with LLM-based component localization~\cite{coca}. CLEW builds on log-to-code mapping to provide explicit investigation context for an agent, rank candidate functions through agent-driven source exploration, and return inspectable links between log events and code.
 
\textbf{LLM-based issue localization.} Agentless narrows the search hierarchically~\cite{agentless}, and agent frameworks support tool-based repository navigation~\cite{yang2024sweagent,zhang2024autocoderover,wang2025openhands}. Several approaches add code structure, such as repository graphs~\cite{ouyang2025repograph}, the heterogeneous code graphs of LocAgent~\cite{locagent}, and call graphs or graph-guided reasoning~\cite{jiang2025cosil,liu2026graphlocator}; others improve exploration scheduling or tool use~\cite{yu2025orcaloca,ma2026tooltrain}. CLEW also uses tools and code structure, supplying the agent with explicit log-to-code anchors organized into diagnostic states and returning evidence links with its function-level ranking.
 
\section{Threats to Validity}\label{sec:threats}

\textbf{Construct Validity.}
On LogLoc-Issues, the reference functions are non-test library functions modified by the fix and present at the base commit. These functions may include changes beyond the root cause. Using the same reference functions across methods supports consistent scoring, but the label definition may still affect absolute scores and relative comparisons, particularly Recall@5. Exact matching of function identities may also count related but non-matching locations as misses. For explanations, the automatic checks assess structural links and associations with log events. They do not establish whether a mechanism path describes the actual cause or whether the explanation helps engineers investigate the failure.

\textbf{Internal Validity.}
LogLoc-Airflow uses injected faults and author-written symptom descriptions, which may influence localization difficulty. Descriptions were written before inspecting the runs and do not explicitly identify the injection location or disclose mutation details. Quoted error messages may retain exception types and code identifiers appearing in the logs. Identifiers that reveal injection points were also anonymized before being provided to any method. These measures reduce direct clues to the fault location but do not eliminate potential bias. One symptom rule was narrowed to the affected task after the second workload configuration was run (Section~\ref{sec:benchmark}); this changed no first-configuration outcome, and the adjustment is given in S4. The compared methods also differ in input length limits, prompts, interaction budgets, and the presentation of log evidence (S5); on LogLoc-Airflow, the baselines read a truncated raw log excerpt, whereas CLEW processes the full log bundle. These differences may contribute to the observed differences in accuracy, cost, and time.

\textbf{External Validity.}
Our evaluation benchmark covers 150 real-world issues from 25 Python repositories and 42 single-function fault-injection cases from 21 injection points in Apache Airflow. The results may not generalize to other programming languages or industrial deployments, and the Airflow results may not extend to faults involving multiple functions. LogLoc-Issues contains reporter-provided log excerpts, whereas LogLoc-Airflow uses a controlled deployment with synthetic workloads. LogLoc-Issues was selected from reports containing log or error evidence, so the results do not establish performance on reports without usable evidence. We examine one perturbation intensity per condition and two base models. Other log formats, missing-evidence patterns, workloads, and models may produce different results. The limited sample sizes also constrain the precision of comparisons, and the stratified results are descriptive.

\section{Conclusion}\label{sec:conclusion}
This paper presented CLEW, an agentic approach for log-guided root-cause code localization, and LogLoc, a benchmark of 192 cases. CLEW combines log-to-code anchoring with LLM-guided backward exploration and returns ranked functions with a proposed mechanism path and associated log observations. CLEW records the highest observed Acc@1 on LogLoc-Issues and matches the leading baselines' Acc@1 on LogLoc-Airflow, with approximately 73\% lower average wall-clock time than LocAgent across all 192 cases at a similar inference cost. The ablations do not establish structured anchoring as a consistent accuracy improvement over the agent's own exploration; anchoring instead provides inspectable links between log evidence and code, whose  structural consistency was assessed. Future work could broaden the industrial evaluation and examine the correctness and practical usefulness of the proposed explanations.

\section*{Data Availability}

The implementation of CLEW and the supplementary material are available~\cite{Clewdata}.

\bibliographystyle{IEEEtran}
\bibliography{references}

\end{document}